\documentclass[5p]{elsarticle}

\usepackage{amssymb}
\usepackage{amsmath}

\usepackage{textgreek}
\usepackage[colorinlistoftodos]{todonotes}
\usepackage[symbol]{footmisc}
\usepackage{float}
\usepackage[hyphens]{url}

\journal{Microelectronic Engineering}

\biboptions{numbers,sort&compress}

\begin{document}

\begin{frontmatter}



\title{Comparison of processing-induced deformations of InP bonded to Si \\ determined by e-beam metrology: direct vs. adhesive bonding} 


\author[1]{Aurimas Sakanas\corref{cor1}}
\ead{ausa@fotonik.dtu.dk}

\author[1]{Elizaveta Semenova}
\author[1,2]{Luisa Ottaviano}
\author[1]{Jesper M{\o}rk}
\author[1]{Kresten Yvind}
\ead{kryv@fotonik.dtu.dk}

\address[1]{DTU Fotonik, Technical University of Denmark, 2800 Kgs. Lyngby, Denmark}
\address[2]{Now with Alight Technologies ApS, 3520 Farum, Denmark}
\cortext[cor1]{Corresponding author}


\begin{abstract}
In this paper, we employ an electron beam writer as metrology tool to investigate distortion of an exposed pattern of alignment marks in heterogeneously bonded InP on silicon. After experimental study of three different bonding and processing configurations which represent typical on-chip photonic device fabrication conditions, the smallest degree of linearly-corrected distortion errors is obtained for the directly bonded wafer, with the alignment marks both formed and measured on the same InP layer side after bonding (equivalent to single-sided processing of the bonded layer). Under these conditions, multilayer exposure alignment accuracy is limited by the InP layer deformation after the initial pattern exposure mainly due to the mechanical wafer clamping in the e-beam cassette. Bonding-induced InP layer deformations dominate in cases of direct and BCB bonding when the alignment marks are formed on one InP wafer side, and measured after bonding and substrate removal from another (equivalent to double-sided processing of the bonded layer). The findings of this paper provide valuable insight into the origin of the multilayer exposure misalignment errors for the bonded III-V on Si wafers, and identify important measures that need to be taken to optimize the fabrication procedures for demonstration of efficient and high-performance on-chip photonic integrated devices.
\end{abstract}


\end{frontmatter}


\section{Introduction}
    
    Availability of advanced silicon processing infrastructure, achieved through the development of integrated electronic circuits, is the main reason for large interest in low cost, high-scale production silicon-based photonic integrated circuits (PICs) \cite{Si-photonics-roadmap}. However, limitations of silicon for realizing efficient on-chip light sources mean that light either has to be coupled into the PIC from an external light source \cite{Si-coupling}, or a different material capable of efficient light emission has to be integrated on silicon \cite{Si-photonics-light-sources}. Perfect candidates for realizing active functionality of the on-chip photonic devices combined with silicon strengths are direct-bandgap group III-V semiconductors \cite{InP-IMOS}.
    
    The main problem for III-V on Si integration is inherent differences in material parameters. Thermal expansion coefficients mismatch is the main limitation for the heterogeneous integration \cite{Direct-bonding-overview,Hitesh-famous,Thin-BCB-bonding}, where two already grown semiconductor wafers are bonded together either directly or through some intermediate layer, such as widely used adhesive benzocyclobutene (BCB) polymer. Specifically, the linear thermal expansion coefficient for InP ($4.6\cdot10^{-6}$ $^{\circ}\text{C}^{-1}$) is almost twice that of Si ($2.6\cdot10^{-6}$ $^{\circ}\text{C}^{-1}$). For the directly bonded InP on Si wafers, the stress in the InP exceeds the critical stress limit and dislocations are generated above 300 $^{\circ}$C temperatures \cite{Pasquariello-critical-stress}.
    
    While critical for the integration, differences between parameters of semiconductors as well as intermediate materials facilitating integration are also very important for further steps of photonic devices fabrication. High temperature treatment is important for oxidation, diffusion processes, contact annealing, and material \mbox{(re-)growth}. After substrate removal, the temperature limit of the critical stress values in III-V directly bonded to Si depends on the III-V layer thickness. An InP layer of 250 nm is able to withstand temperatures encountered during the epitaxial growth without exceeding the critical stress values and generation of dislocations \cite{Fujii-NTT}. BCB bonding is generally not suitable for processing where prolonged high temperature exposure is required or certain chemical resistance limitations apply, as after curing the glass transition occurs above 350 $^{\circ}$C temperature \cite{BCB-datasheet}. In addition, CMOS compatibility may impose stricter temperature requirements.

    Alignment for the multilayer exposure is a standard procedure in the semiconductor device fabrication process, and the level of alignment accuracy needed for the advanced photonic integrated devices is very high. This may be for embedding gain material into the photonic crystal cavities and waveguides for lasers \cite{NTT-PhC-laser}, modulators \cite{NTT-PhC-modulator} and photodetectors \cite{NTT-PhC-photodetector}, contacts for electrical operation of such devices \cite{Lupi-PhC-pillar}, light coupling to/from active III-V layer into passive Si circuitry \cite{InP-Si-Ebeam-alignment}, or even combinations of these. Many factors can, through the complex wafer geometry changes (arising from deposition of films with non-uniform residual stress or bonding \cite{InP-Si-XRD-strain-maps}, in-plane stretching deformations induced by wafer chucking during lithography \cite{Overlay-wafer-chucking}), alter initially defined device patterns and their relative positions to alignment marks during wafer processing in-between exposure steps. In advanced silicon electronics manufacturing, high-resolution wafer geometry measurements combined with mechanics-based models are used as standard process control methods of predicting and correcting processing-induced overlay errors between lithography steps \cite{overlay-silicon} to meet the ever-tightening error budgets. Unfortunately, implementation of such methods in practice is a complicated task and may not apply for less-mature Si photonics processing, thus semi-automatic (or even manual) alignment in less-modern DUV or electron beam lithography systems is generally used \cite{InP-Si-Ebeam-alignment}.
    
    Wafer bonding adds additional level of complexity in the process, and affects the alignment accuracy depending on the type of bonding used \cite{alignment-table}. Unless some special self-aligning device designs \cite{PhC-self-aligned} or pattern distortion compensation methods are adapted, integration precision usually only relies on a high density of alignment marks near device structures. Better understanding of the wafer-scale non-uniformities of the bonded III-V on Si wafers is therefore needed for the design and processing optimization.
    
    In this paper, we present e-beam metrology measurements of processing-induced wafer distortion of 250 nm InP layers bonded to 2\texttt{"} Si substrates by adhesive BCB polymer and direct bonding techniques. To our knowledge, this is the first systematic experimental attempt to quantitatively estimate degree of misalignment arising on the wafer-scale for the heterogeneously bonded III-V layers on Si with such precision.

\section{Materials and Methods}
\label{methods}

	In this section, the process of wafer bonding using adhesive BCB polymer as well as the direct bonding is described. The formation of alignment marks in the bonded InP layers in a single- and double-sided processing configurations is explained. Finally, the main principles of using e-beam as metrological tool to quantify processing-induced pattern deformation is introduced and its precision is assessed. Process configurations and methods represent real conditions encountered during photonic device fabrication, and is not devised to illustrate the best-case scenario.

\paragraph{Wafer bonding and formation of alignment marks}

	Multiple wafers have been prepared in a MOCVD reactor by epitaxially growing a 250 nm InP layer on top of a lattice-matched etch-stop InGaAs layer on a 2\texttt{"} (100)-orientation InP substrates. These wafers were then divided into two groups for the BCB and direct bonding experiments.
    
    Prior to the BCB bonding, dry etching was used to define the alignment marks in the top epitaxial layer on InP wafer. For this, ZEP520A positive e-beam resist of $\sim$500 nm was spin-coated onto the PECVD-deposited non-stoichiometric 200 nm silicon nitride (SiNx) etch mask. An array of 21x21 alignment marks (crosses, with an arm length of 375 \textmu m and a width of 2 \textmu m) was exposed in e-beam with a distance of 1.5 mm between the centres of two neighbouring marks, defining an array pattern of 3x3 cm. After development, the pattern was transferred to the underlying SiNx layer using RIE etching with CHF$_3$/O$_2$, and to the 250 nm InP layer by a chlorine-based ICP etch (the openings are etched into the InP layer instead of having isolated material as the alignment marks, because in this way measurements are not affected by possible localized distortions or adhesion issues of the marks). The rest of the SiNx was removed by another RIE etch. Finally, a 1 \textmu m thick SiO$_2$ layer was deposited on the InP wafer in PECVD, while a 1.3 \textmu m thick SiO$_2$ layer was thermally grown on a 2\texttt{"} (100)-orientation Si wafer. The choice of introducing SiO$_2$ on both wafers is partially related to increased adhesion to the BCB, but it also serves a purpose in device operation \cite{Xue-PhC} and optical-coupling schemes \cite{InP-Si-Ebeam-alignment}.
    
	Proceeding with the BCB bonding, both InP with the defined alignment marks and Si wafers were treated with AP3000 adhesion promoter, and then a $\sim$2.5 \textmu m thick layer of BCB (Cyclotene 3022-46) was spin-coated on the Si wafer, followed by 5 min soft-bake at 90 $^{\circ}$C. Stacked on top facing each other wafers were placed inside a wafer bonder to cure/polymerize the BCB for 1 hour at 250 $^{\circ}$C temperature in vacuum under an applied force of 2 kN. After the bonding was complete, the InP substrate was removed by wet etching in HCl, and the etch-stop InGaAs layer was removed in H$_2$SO$_4$:H$_2$O$_2$:H$_2$O 1:8:8 solution.

\begin{figure}
\centering
\includegraphics[width=8cm]{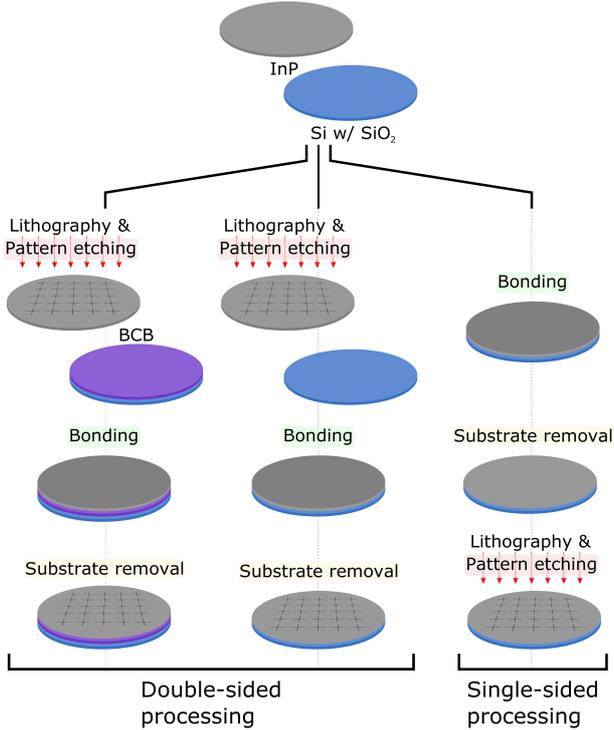}%
\caption{\label{fig:1}Schematic representation of double-sided and single-sided processing methods used for metrology experiments.}%
\end{figure}

    The wafer processing and bonding flow described previously can be considered as an example of the double-sided process, in which the initial structures are first formed on the top-side of the III-V layer, followed by wafer bonding, and subsequent alignment and processing from the back-side of the III-V layer after uncovering it by the substrate/sacrificial layer removal. If wafer bonding precedes the formation of the initial structures, the alignment and further processing are realized from the same side of the III-V layer, hence denoted as the single-sided process. These methods are schematically illustrated in Fig. \ref{fig:1}.
    
    Both single- and double-sided processing methods were used to prepare a batch of directly bonded wafers for the metrology experiments. The process flow for fabricating the alignment marks in the top InP layer is very similar to the one presented earlier, with two main exceptions: 1) for transferring the alignment marks from the SiNx mask to the InP layer with reduced sidewall roughness, we used anisotropic crystallographic-orientation dependent H$_3$PO$_4$:HCl solution with 1:4 ratio instead of ICP etching, and 2) no PECVD SiO$_2$ layer was deposited on the InP side before the wafer bonding, as no thick adhesive layer separates the InP from the thermal glass on silicon in the final stack.
    
    The direct InP to Si wafer bonding was performed using an intermediate thin Al$_2$O$_3$ layer \cite{Hitesh-famous} to achieve a high bond-strength and avoid problems associated with bonding of two dissimilar surfaces. Both InP and Si wafers were placed in an ALD chamber, where a thin 2 nm Al$_2$O$_3$ layer was deposited at 200 $^{\circ}$C on each side of the wafers. Al$_2$O$_3$ provides surfaces rich in OH-groups, which results in the immediate hydrogen bonding facilitated by the Van der Vaals forces after placing the wafers together \cite{Bonding-main}. The surfaces were further bound and the bond strength was increased by placing the wafer stack in a wafer bonder for 1 h at 300 $^{\circ}$C in vacuum with a force of 2 kN applied to the stack. Next, the InP substrate and InGaAs etch-stop layer was removed as described previously leaving the 250 nm InP layer on Si.
    
    From the fabrication standpoint, the single-sided processing approach is advantageous as the crucial bonding step is performed early in the III-V processing stage. This reduces the chance of introducing particles on the wafer surface, which cause the decrease of bonding yield and in the extreme case can even lead to bonding failure. In terms of wafer surface preparation prior to bonding, no special cleaning was done neither for the InP wafers taken directly from the MOCVD reactor after the epitaxial growth, nor for the InP wafers with the pre-etched alignment marks. While in the first case it was possible to obtain defect-free bonding, preventing voids on the pre-processed wafers after the InP substrate removal is a much more difficult task, which in principle is achievable with proper cleaning and/or chemical-mechanical polishing procedures.
    
\paragraph{E-beam metrology measurements procedure}
    
    In order to determine processing-induced deformation for different InP to Si wafer bonding and processing approaches presented in this paper, we used the JEOL JBX-9500FSZ \mbox{e-beam} system both for lithography and as a metrological SEM. Quality control of the overlay accuracy for standard Si processing demonstrated the achievable performance to be well within the specifications of the machine \cite{Our-Ebeam}.
    
    For pattern writing, 2\texttt{"} wafers were clamped in the central slot of a titanium cassette, loaded onto the e-beam stage, and thermally stabilized for a few hours to limit temperature drift to within 0.01 $^{\circ}$C during the single scan. Thermal drift is the most dominant overlay error source \cite{Ebeam-error-sources}, thus using Ti cassette allows for better thermal stability, as its thermal expansion coefficient is a few times smaller than that of aluminium. Next, the instrument column was calibrated for an electron beam current of 6 nA, which in conjunction with an aperture size of 80 \textmu m corresponds to a beam diameter just below 10 nm.
    
    The e-beam metrology procedure consists of three main parts: sample positioning using any two alignment marks, measurement of an array of alignment marks, comparison and correction of designed and measured positions.
    
    Sample positioning is realized by placing the bonded InP-on-Si wafer into the cassette, manually adjusting it in the cassette window, and then tightly clamping from the back-side with the leaf-spring loaded back-plate. The cassette is then loaded onto the e-beam stage, where a 100 keV beam scans across arms of two (global) alignment marks in \textit{x}- and \textit{y}-directions, resulting in a detectable backscattered-electrons signal from the topography and Z (atomic number) contrast, which in turn allows to determine the centre-positions of the scanned marks. Measured centre-positions are then compared to designed values, and wafer magnification (gain), shift and rotation are corrected by the machine for subsequent array scanning.

    Metrology scanning is performed right after the e-beam calibration and wafer position correction. A metrology scan file contains the design coordinates of an array of alignment marks and parameters for optimal signal detection. Six independent scans are performed on each wafer in total, where each mark in the array is scanned at three different arm-positions and each measurement is performed twice by changing the stage movement pattern (\textit{x}- and \textit{y}-raster type scans) with a fine-scan width of 12 \textmu m or less. Scanning at different positions allows to obtain a complete set of data even if some of the experimental values are corrupt due to, for example, particles or bonding defects at parts of the alignment mark. Also, averaging over multiple measurements reduces random position detection errors.
    
    Finally, a data file is generated for each array scan in which the difference between the designed and measured position of each individual alignment mark is recorded. After removing any possibly corrupted data points by erroneous detection, the measurements are averaged and vector maps are plotted after calculation and correction of the linear components of magnification $M_x$ and $M_y$, shift and rotation \cite{Overlay-analysis}. In this manner, each and every vector represents the direction and magnitude of deviation in \textit{x}- and \textit{y}-directions from the designed alignment mark position.
    
\paragraph{Quantifying baseline accuracy for e-beam metrology}

	Before analysing e-beam metrology results for the bonded wafers, it is important to estimate accuracy and limitations of the metrology system itself. For this purpose, we used a bare 2\texttt{"} Si wafer. Following exposure and dry etching procedure of alignment marks similar to that described in section \ref{methods}, the wafer was placed back into the e-beam cassette to estimate residual wafer-scale errors.

\begin{figure}
\centering
\includegraphics[width=8cm]{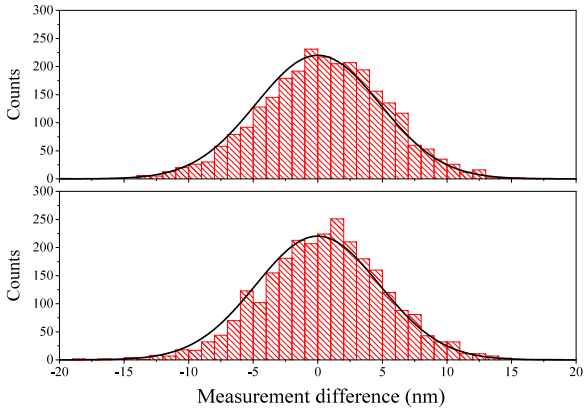}%
\caption{\label{fig:2}Histograms representing measured differences between repeated metrology scans for (top) \textit{x}- and (bottom) \textit{y}-coordinates of alignment marks in the 21x21 size array defined on the Si wafer. Solid lines are fitted normal distribution curves.}%
\end{figure}

    Overall uniformity and repeatability of measurements is assessed by subtracting vector maps obtained after placing and re-placing Si wafer in the cassette slot. The data is plotted in a form of histogram for \textit{x}- and \textit{y}-coordinates separately (Fig. \ref{fig:2}), and the statistical 3\textsigma{} standard deviation is found to be 15 nm for both.

\section{Results}
	
    In this section, processing-induced alignment marks pattern deformation is quantified using the e-beam metrology measurement results for the BCB and directly bonded wafers processed in the double- and single-sided manner. After discussion of the observed characteristic features in each case, a comparison is made.

\paragraph{Distortion of BCB bonded double-side processed wafers}
	
\begin{figure}
\centering
\includegraphics[width=8cm]{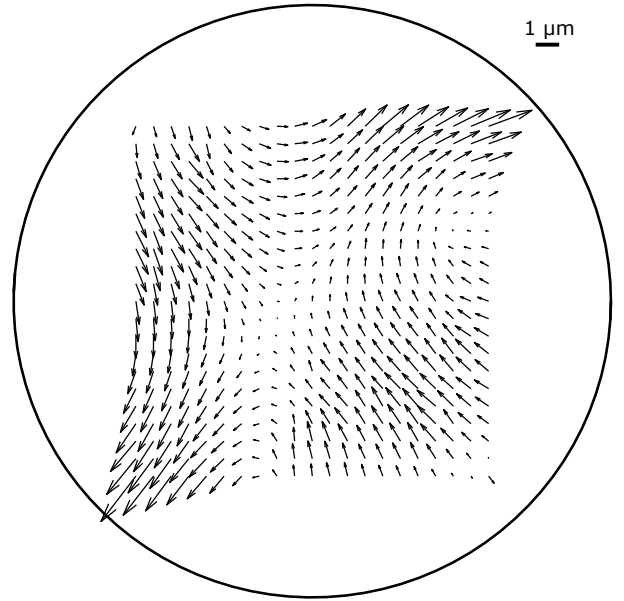}%
\caption{\label{fig:6}Vector map representing processing-induced deformation of the double-side processed InP layer BCB bonded to 2\texttt{"} Si wafer. Black circle indicates outline of the 2\texttt{"} wafer; the scale bar is for the vectors. A uniform linear expansion ($M_x-1=1.4\cdot10^{-4}$, $M_y-1=1.6\cdot10^{-4}$) has been subtracted from the measured data, pattern shift and rotation are corrected.}%
\end{figure}

	As discussed in section \ref{methods}, for the BCB bonded wafers the array of alignment marks is scanned from the side opposite to which it was defined, representing the double-sided processing approach. The 3\textsigma{} standard deviation for the detection of dry etched alignment marks in InP was estimated to be below 40 nm for \textit{x}- and \textit{y}-directions, worse than for Si. Nevertheless, the pattern distortion determined after the correction of magnification, shift and rotation is found to be much greater than that (Fig. \ref{fig:6}). In fact, in large parts of the wafer away from the centre, the measured distortion is on the order of 1 \textmu m and in some parts exceeds even 2 \textmu m. It seems that rather thick 2.5 \textmu m BCB layer allows for stress relaxation between InP and Si wafers, and after the InP substrate and sacrificial layer removal, the 250 nm InP layer is found to be predominantly expanded along one diagonal direction, while contracted along another (implying the presence of an orthogonal distortion component \cite{Overlay-analysis}). Although linear, orthogonality correction is beyond the standard e-beam correction capabilities, meaning that the pattern distortion in Fig. \ref{fig:6} represents the typical situation after correction for the multilayer exposure in device fabrication. Repeated measurements on the BCB bonded wafers show qualitatively similar orthogonal distortion signatures, but of different orientations. This  indicates that such BCB bonding-induced distortion is generic, but with randomly oriented orthogonality.
    
    It was previously reported \cite{Niklaus-BCB} that partially curing BCB prior to wafer bonding can prevent polymer reflow and to some degree improve the alignment between two Si wafers. While such optimization prior to wafer bonding, using thinner BCB layer, and temperature compensation of the thermal expansion coefficients differences for InP and Si (by adjusting top and bottom chuck temperatures in the wafer bonder) could potentially reduce pattern distortion, this is a matter of a whole separate study.
    
	The consequences of this significant non-uniform deformation observed for the double-side processed BCB bonded wafer is that proper wafer-scale correction by using alignment marks with larger-spacing is very difficult, and achieving high-accuracy alignment to the initial exposure on the InP wafer after the BCB bonding is difficult. A high density of closely-placed alignment marks would be needed for finer compensation, however this would result in drastically prolonged pattern writing time, increased design complexity and added overall fabrication cost.

\paragraph{Distortion of directly bonded double- and single-side processed wafers}

\begin{figure}
\centering
\includegraphics[width=8cm]{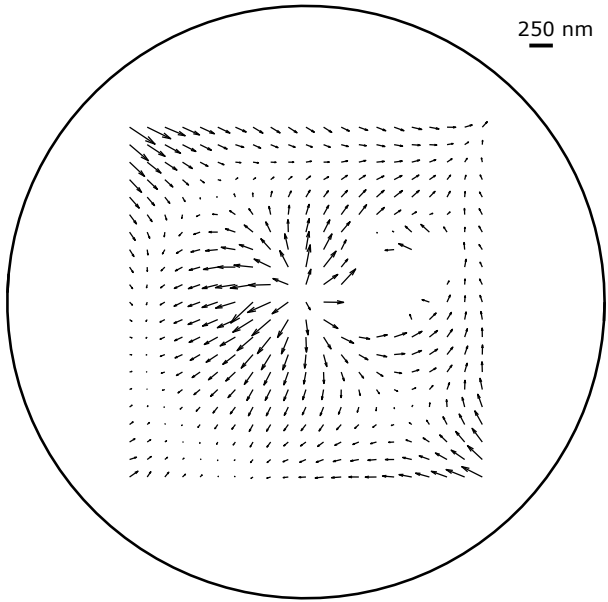}%
\caption{\label{fig:8}Vector map representing processing-induced deformation of the double-side processed InP layer directly bonded to 2\texttt{"} Si wafer. Empty vector map regions correspond to the bonding defects. Black circle indicates outline of the 2\texttt{"} wafer; the scale bar is for the vectors. A uniform linear expansion ($M_x-1=3.4\cdot10^{-5}$, $M_y-1=2.2\cdot10^{-5}$) has been subtracted from the measured data, pattern shift and rotation are corrected.}%
\end{figure}

	For the direct bonding, both double- and single-sided processing approaches were utilized. Double-sided processing adds additional requirements for wafer handling to avoid particle deposition on the surface before bonding. While a thick BCB layer can embed some of the smaller particles and prevent the formation of bonding defects in the InP layer, the thin Al$_2$O$_3$ used as an intermediate layer for the direct bonding is incapable of that. The InP wafers in these experiments are subject to multiple processing steps during the formation of the alignment marks. Thus, the appearance of bonding defects is expected, and indeed is seen in Fig. \ref{fig:8} as a void in the vector map. Obtaining high bonding yield is challenging for any kind of direct bonding, however the use of the single-sided processing approach reduces the number of wafer handling steps prior to bonding, since the InP wafer can be bonded to the Si substrate right after the epitaxial growth and Al$_2$O$_3$ deposition in ALD.
 
    Wet instead of dry etching of the alignment marks was used for the directly bonded wafers to reduce sidewall roughness and avoid high temperature during transfer, however only minor improvement was determined, with 3\textsigma{} below 35 nm for \textit{x}- and \textit{y}-directions. The metrology measurements of the directly bonded double-sided processed wafer shows misalignment errors on the order of a few hundred nm's across the wafer (Fig. \ref{fig:8}). Disproportional expansion between the centre and outer parts might be caused by the improper pre-bonding if the initial bonding did not originate from a single-site, but rather multiple locations \cite{Bonding-main}. Deformation asymmetry around the bonding defect region indicates its local influence.
    
    Significant improvement of the direct bonding over the BCB bonding is apparent, which would allow for reasonable alignment precision in most parts of the wafer even with the use of distantly-spaced alignment marks.
    
\begin{figure}
\centering
\includegraphics[width=8cm]{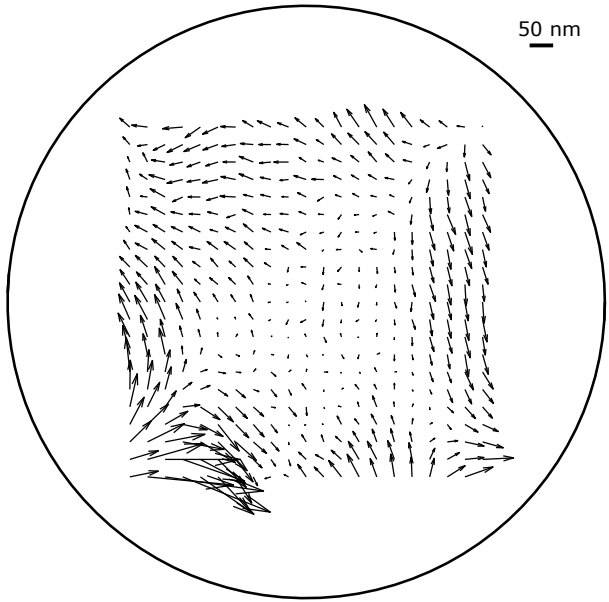}%
\caption{\label{fig:9}Vector map representing processing-induced deformation of the single-side processed InP layer directly bonded to 2\texttt{"} Si wafer. Black circle indicates outline of the 2\texttt{"} wafer; the scale bar is for the vectors. A uniform linear expansion ($M_x-1=5\cdot10^{-6}$, $M_y-1=3.7\cdot10^{-6}$) has been subtracted from the measured data, pattern shift and rotation are corrected.}%
\end{figure}

    Finally, the pattern distortion for the directly bonded wafer using the approach of single-sided processing is shown in Fig. \ref{fig:9}. After linear data correction, the central part of the wafer has high uniformity with the residual distortion errors below 50 nm. However, larger deformations are observed in the corners, especially the lower left. The pattern exposure, transfer and metrology for the single-sided processing are performed after the bonding, thus it is not expected that these deformations are bonding-induced. Rather, simple calculations similar to those in \cite{Overlay-wafer-chucking} indicate that strain on the order of $5\cdot10^{-6}$ can be easily induced on the pattern surface by the deposited stressed SiNx layer (for our PECVD SiNx $\sim$450 MPa), which due to the wafer clamping can result in the in-plane pattern distortion on the order of 100 nm. At this scale, additional distortion components contribute, such as localized stress variations (particles on the wafer surface when clamping in the e-beam cassette, deposited film non-uniformity) and/or electromagnetic effects near the e-beam cassette slot edges.

	High uniformity of the central part of the wafer allows for precise alignment using even distantly spaced alignment marks compatible with the standard DUV stepper reticle field size, which can increase device throughput maintaining high alignment accuracy between exposures.
    
\begin{figure}
\centering
\includegraphics[width=8cm]{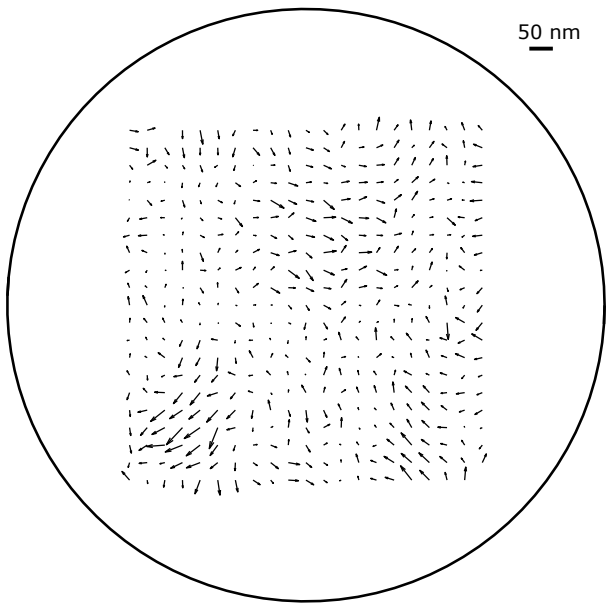}%
\caption{\label{fig:10}Difference-vector map obtained by subtracting metrology measurements of the pattern on the InP layer directly bonded to 2\texttt{"} Si wafer before and after MOVPE anneal. Black circle indicates outline of the 2\texttt{"} wafer; the scale bar is for the vectors.}%
\end{figure}

    Finally, to assess the wafer stability at high temperatures, we performed annealing two times at 650 $^{\circ}$C for 15 min in the phosphine atmosphere, followed by the metrology after each annealing step. The difference-vector map between the initially measured wafer, and after the second annealing (Fig. \ref{fig:10}) indicates that high temperature processing for even extended durations does not introduce considerable additional InP layer deformations. The surface quality of the wafer also does not seem to change, although we observed some degradation of the alignment marks, which reduced the metrology precision. Some kind of protection might be needed if they are already fully defined in the InP layer prior to the annealing step.
    
\paragraph{Comparison of deformations in bonded wafers} 

	When comparing the adhesive-BCB and directly bonded InP to Si wafers (and inherently double- and single-sided type processing), several features are observed.
    
    The pattern defined by the alignment marks in the InP layer is distorted after the BCB bonding non-uniformly and with a significantly greater magnitude than for the directly bonded wafer using the double-sided processing approach. It means that to achieve the same level of multilayer exposure alignment accuracy, considerably denser array of alignment marks would be needed for the BCB bonding, complicating the design, increasing the exposure as well as the alignment time, and subsequently the cost. Furthermore, processing temperatures are limited by the BCB polymer. On the other hand, direct wafer bonding is more demanding than BCB bonding. It requires greater care when dealing with particles, including wafer handing, chemical cleaning and polishing, thus the high bonding yield is in principle harder to achieve.
    
	The highest wafer-scale alignment precision is achievable on the directly bonded and single-side processed wafer with relaxed requirements for the density of alignment marks. Wafer cleanliness requirements are relaxed, as is only critical prior to bonding. Subsequent wafer processing at high temperatures is also tolerable.
    
\section{Conclusions}

	An e-beam metrology technique was employed to study processing-induced deformation of the exposed pattern in 2\texttt{"} InP wafers bonded to Si substrates of the same size processed in three different ways. The conclusions are twofold: first, bonding-induced pattern deformation is dominant for the double-sided processing approach, where the initial processing is performed from one InP wafer side, followed by bonding and substrate removal, revealing another InP-side from which the alignment is then performed. When using thick adhesive BCB bonding, deformation on the order of 1 \textmu m is observed across the entire 2\texttt{"} wafer. It can be significantly reduced by using direct wafer bonding instead, however requirements for surface cleanliness are largely increased in this case to achieve high bonding yield after wafer processing. Second, the smallest deformation was experimentally determined for the direct wafer bonding combined with the single-sided processing approach, in which case the processing and alignment is performed after the InP wafer is bonded to Si. At this scale, the alignment accuracy approaches the case when a bare InP or Si wafer is used and is limited by the pattern deformation induced during wafer processing by the wafer clamping in the e-beam cassette. This holds even after prolonged treatment at 650 $^{\circ}$C temperature.
    
    Finally, it should be noted that the metrology method described in this paper could be customized and used as a convenient tool to determine processing-induced pattern deformation during device fabrication. This knowledge could be applied when accounting for and correcting fine residual-errors prior to the overlay exposure(s).


%
%

%

\section*{Acknowledgments}
The authors acknowledge financial support from Villum Fonden via the NATEC (Nanophotonics for Terabit Communications) Centre (grant no. 8692). Alexandra Lupi is acknowledged for performing initial measurements prior to this study; Peixiong Shi and Thomas Aar{\o}e Anh{\o}j from Danchip for setting-up e-beam for metrology and fruitful discussions. Fabrication and all experiments were performed in the Danchip cleanroom at DTU.

\bibliography{ReferencesMendeley}

\end{document}